\documentclass[11pt]{article}
\begin{document}
\title{Nonstandard introduction to squeezing of the electromagnetic field}
\author{Iwo Bialynicki-Birula\\Center for Theoretical Physics PAN and
College of Science\\Lotnik\'ow 46, 02-668 Warsaw, Poland
\thanks{e-mail: birula@cft.edu.pl}
\thanks{Presented at the XXXVIII Cracow School of Theoretical Physics, Zakopane, June 1-10, 1998. To be published in Acta Physica Polonica B.}}
\maketitle

\newcommand{\lnabla}{\!\stackrel{\leftarrow}{\nabla}\!}
\newcommand{\lpt}{\!\stackrel{\leftarrow}{\partial}\!}
\newcommand{\llapl}{\!\stackrel{\leftarrow}{\Delta}\!}

\begin{abstract}
This article contains a review of an alternative theory of squeezing, based
entirely on the wave function description of the squeezed states. Quantum field
theoretic approach is used to describe the squeezing of the electromagnetic
field in its most complete form that takes into account temporal and spatial
characteristics of the squeezed state. An analog of the Wigner function for the
full electromagnetic field is introduced and expressed in terms of second order
correlation functions. The field-theoretic approach enables one to study the
propagation of the ``squeezing wave'' in space-time. A simple example of weak
squeezing, that allows for all calculations to be done analytically, is
discussed in detail.
\end{abstract}

\section{Introduction}

The standard description of squeezing of the electromagnetic field
\cite{sto,wm,mw,scully} employs the annihilation and creation operators
associated with the mode decomposition of the field operators. Subsequently, only
one or two modes are singled out for the analysis of the squeezing phenomena.
This procedure gives a very effective tool to explain most squeezing
experiments. However, it will not be adequate if one attempts to describe the
propagation of squeezing in space-time, when many modes are participating in the
process and the state of the field exhibits strong correlations between
different modes. In order to accommodate such situations, I shall introduce a
description of squeezing, started in Refs. \cite{ibb1,ibb2,bbbb} and continued
more recently \cite{ibb3}, that employs formal methods of quantum field theory
and enables one to produce compact and closed expressions for the space-time
characteristics of the squeezing phenomena. Very convenient tools in this
analysis will be the wave functional and the Wigner functional of the quantized
electromagnetic field. These objects require a specification of the reference
frame since they depend on the field 3-vectors ${\bf B}$ and ${\bf D}$. At a
first glance such a description may look noncovariant and the author of Ref.
\cite{burgess} has fallen into this trap. However, as I show in Sec.
\ref{wig_rel}, despite the absence of tensorial indices, our formulation is
fully relativistically covariant.

Since the electromagnetic field may be viewed as one grand, multidimensional
harmonic oscillator, many properties of squeezed states can be clearly seen in a
simpler case of an $N$-dimensional harmonic oscillator in quantum mechanics.
Using this example, I introduce in Sec. \ref{multi} the necessary formalism and
I describe in Sec. \ref{ur} special properties of the squeezed states related to
uncertainty relations for positions and momenta. Next, I solve in Sec. \ref{time_evol}
the time evolution problem for squeezed states and I introduce in Sec. \ref{wigner}
the description of squeezed states in terms of the Wigner function.

In Sec. \ref{qed}, the formalism developed for squeezed states of the quantum
mechanical oscillator will be carried over to quantum electrodynamics. Even
though the electromagnetic field corresponds to an {\em infinite-dimensional}
oscillator, there is a complete formal correspondence between these two cases.
This correspondence enables one to write complete expressions for the relevant
quantities in quantum electrodynamics in a closed form even though the rigorous
mathematical foundation for this theory is often lacking. I address In Sec. \ref{prop}
the problem of the propagation of squeezing of the electromagnetic field
in space-time and I discuss in detail simple example of a small perturbation.
Finally, I introduce in Sec. \ref{wig_rel} the Wigner functional for the
full quantized electromagnetic field --- an analog of the quantum-mechanical
Wigner function and I discuss its relativistic properties.

There is no universally adopted terminology concerning the squeezed states and
some authors extend the notion of squeezed states to almost any state that
exhibits some kind of squeezing. Throughout this paper I shall use the term
squeezed states in a narrow sense to denote the states whose wave function is a
Gaussian. Such states, as discussed in Sec. \ref{ur}, saturate various forms of
the uncertainty relations for positions and momenta. They also can be obtained,
as is discussed in the standard theory of squeezing, from the vacuum state by
the action of operators that are quadratic in the annihilation and creation
operators in the exponent \cite{sto,wm,mw,scully}.

\section{Multidimensional harmonic oscillator\\ in quantum mechanics\label{multi}}

Squeezed states are usually defined in terms of annihilation and creation
operators, but the description in terms of the wave function in coordinate
representation or in momentum representation, which I have adopted here, is very
useful for a general analysis:
\begin{itemize} {\item It offers a geometric picture of the squeezing phenomenon that is easy to grasp intuitively}
{\item It leads to compact formulas that exhibit the correlations between
various oscillators} {\item It enables one to give a simple general description
of time evolution} {\item It uses the smallest possible number of parameters to
describe an arbitrary squeezed state}
\end{itemize}
Of course, the formalism based on annihilation and creation operators is often
very convenient but for the uniformity of my exposition I shall restrict myself
solely to the description in terms of wave functions.

The Hamiltonian of an $N$-dimensional harmonic oscillator will be taken in the
form (summation convention for repeated indices will be used throughout)
\begin{eqnarray}
 H = \frac{1}{2}\bigl(g^{ij}p_ip_j + u_{ij}x^ix^j\bigr),\label{hamiltonian}
\end{eqnarray}
where both matrices $g$ and $u$ are symmetric and positive definite. One might
also include the quadratic terms that mix $x$ and $p$ but I shall omit them
since they are not needed in the analysis of the electromagnetic field. Changing
the variables, one may transform $g^{ij}$ to a multiple of the Kronecker delta.
\begin{eqnarray}
g^{ij} = \frac{1}{m}\delta^{ij}.
\end{eqnarray}
By an additional orthogonal transformation, one may bring the potential matrix
$u$ to a diagonal form. The eigenvalues $u_i$ of this matrix are multiples of
the squares of the characteristic frequencies $\Omega_i$ of the oscillators,
$u_i = m \Omega^2_i$.

The ground state of the quantum-mechanical system with the Hamiltonian
(\ref{hamiltonian}) can be obtained by simply plugging in a Gaussian wave
function of the form
\begin{eqnarray}
 \psi_0({\bf r}) = ({\rm Det}\,\Omega)^{1/4}(m/\pi\hbar)^{N/4}
 \exp(-\frac{m}{2\hbar}x^i\Omega_{ij}x^j),\label{ground}
\end{eqnarray}
into the time-independent Schr\"odinger equation $H\psi = E\psi$ and by solving
the matrix equation
\begin{eqnarray}
 m^2 \Omega g \Omega = u.\label{ground_eq}
\end{eqnarray}
The solution for $\Omega$ must be chosen as a symmetric and positive matrix. In
our diagonal representation $\Omega$ has the form
\begin{eqnarray}
 \Omega = \left( \begin{array}{ccccc} \Omega_1 & 0 & 0 & \cdots & 0\\ 0 &
 \Omega_2 & 0 & \cdots & 0\\ \cdot&\cdot&\cdot&\cdot&\cdot\\
 \cdot&\cdot&\cdot&\cdot&\cdot\\ \cdot&\cdot&\cdot&\cdot&\cdot\\ 0 & 0 & 0 &
 \cdots &\Omega_n\end{array}
\right).
\end{eqnarray}
In addition to the ground state, there exist other Gaussian wave functions. The
most general normalized Gaussian wave function of an $N$-dimensional harmonic
oscillator is characterized by two real vectors $\xi^i $ and $\pi_i $ and two
real symmetric $N\times N$ matrices forming one complex matrix $K_{ij} = a_{ij}
+ ib_{ij}$,
\begin{eqnarray}
 \psi({\bf r}) &=& ({\rm
 Det}\,a)^{1/4}(m/\pi\hbar)^{N/4}\exp(-i\xi^i\pi_i/2\hbar) \nonumber \\
 &\times&\exp\bigl[-\frac{m}{2\hbar}(x^i - \xi^i)K_{ij}(x^j - \xi^j) + i\pi_i
 x^i/\hbar\bigr],\label{gauss}
\end{eqnarray}
where the phase factor has been introduced to maintain a symmetry between the
position (\ref{gauss}) and the momentum representation (\ref{gauss1}).

Gaussian wave functions are special: the Gaussian form of the initial wave
function of a harmonic oscillator remains Gaussian during the whole time
evolution. This property has been recognized right after the discovery of wave
mechanics \cite{ken}. In addition, as shown in Sec. \ref{ur}, the Gaussian wave
functions saturate the multidimensional generalization of the
Schr\"odinger-Robertson uncertainty relations \cite{schr,robert} and the
entropic uncertainty relations \cite{bbm}. Gaussian distributions in probability
theory are also exceptional: all higher order correlation functions can be
expressed in terms of the lowest correlation functions. This property is
reflected in the properties of the Wigner function for the squeezed states, as
shown in Sec. \ref{wigner}.

The notion of a squeezed state always refers to a specified Hamiltonian. Given a
Hamiltonian of the form (\ref{hamiltonian}) one can find the ground state but
the inverse problem does not have a unique solution because for every symmetric
and positive matrix $\Omega$ there is a whole family of Hamiltonians of the form
(\ref{hamiltonian}) such that the ground state is described by (\ref{ground}).
The only condition on the matrices $g$ and $u$ is that they satisfy the
condition (\ref{ground_eq}). One can even find a Hamiltonian for which the
ground state is described by a complex matrix $K$ but if the imaginary part of
$K$ does not vanish the Hamiltonian must also contain the mixed term
$f_i^{\,j}x^i p_j$.

The choice of the ground state provides us with a standard against which one can
measure the departures of the mean square deviations in positions and momenta
from their {\em normal} (i.e. ground state) values. This enables one to define
the notion of squeezing. In particular, it enables one to relate the scale in
the momentum variables to that of the position variables. All Gaussian wave
functions correspond to squeezed states, except the one describing the ground
state. When $a$ is ``larger'' than $\Omega$, the state is squeezed in the
position variables and when it is smaller, it is squeezed in the momentum
variables. Of course, since $K$ is a matrix, the degree of squeezing may depend
on the direction.

The energy of the system separates into the center of mass part and the internal
motion part
\begin{eqnarray}
 E &=& E_{\rm cm} + E_{\rm int},\label{energy}\\ E_{\rm cm} &=&
 \frac{1}{2}g^{ij}\pi_i\pi_j + \frac{1}{2}u_{ij}\xi^i \xi^j,\label{center}\\
 E_{\rm int} &=& \frac{\hbar}{4}{\rm Tr}\{m g b a^{-1} b + m g a + u
 a^{-1}/m\}.\label{internal}
\end{eqnarray}
The energy of the internal motion tends to zero in the classical limit, when
$\hbar \to 0$, because the Gaussian shrinks to zero and one is left with the
motion of a classical point-like particle. It is worthwhile to observe that the
Gaussian solutions of the Schr\"odinger equation correspond to two independent
autonomous mechanical systems with their Hamiltonians given by the expressions
(\ref{center}) and (\ref{internal}). The canonical variables for these two
systems, as has been shown in Ref. \cite{ibb1}, in addition to $(\xi^i, \pi_i)$,
are the components of the matrices $a^{-1}$ and $b$.

From the expression for $E_{\rm int}$ one may obtain the following formulas for
the first and second variation of the internal energy around the ground state
\begin{eqnarray}
 \delta E_{\rm int} &=& \frac{\hbar}{4} {\rm Tr}\{\delta a - u a^{-1}\delta a
 a^{-1}/m\},\label{eq}\\
\delta^2 E_{\rm int} &=& \frac{\hbar}{2}{\rm
 Tr}\{\delta b\,\Omega^{-1}\,\delta b + \delta a\,\Omega^{-1}\,\delta
 a\},\label{stab}
\end{eqnarray}
where I have used $g^{ij} = \delta^{ij}/m$. The equilibrium condition derived
from $\delta E_{\rm int} = 0$ ($\delta a$ is arbitrary) coincides with the
equation (\ref{ground_eq}). The quadratic form defining the second variation is
clearly positive confirming the fact that the energy of the ground state is the
lowest one.

The formula (\ref{energy}) can also be used to find the change in the energy of
the ground state due to a change of the matrices $g$ and $u$ describing the
system. Here one can see the advantage of keeping $g$ and $u$ arbitrary.
\begin{eqnarray}
 \delta E_{\rm int} &=& \frac{\hbar}{4}{\rm Tr}\{m \delta g\,a + m g\delta a +
 \delta u\,a^{-1}/m - ua^{-1}\delta aa^{-1}/m\}\nonumber\\ &=&
 \frac{\hbar}{4}{\rm Tr}\{m \delta g\,a + \delta u\,a^{-1}/m\}.
 \label{hell_feyn}
\end{eqnarray}
Since for the ground state the variations $\delta a$ of the matrix $a$ cancel
out, this formula may be viewed as an analog of the Feynman-Hellmann theorem
\cite{hell,feyn}.

Analogous results are obtained in momentum representation. The general
normalized Gaussian wave function in momentum space is obtained by the Fourier
transformation of the position space wave function (\ref{gauss}) and it reads
\begin{eqnarray}
 \phi({\bf p}) &=& ({\rm Det}\,a)^{-1/4}(\pi\hbar m)^{-N/4}
 \exp(i\xi^i\pi_i/2\hbar)\nonumber\\ &\times&\exp\bigl[-\frac{1}{2m\hbar}(p_i -
 \pi_i)(K^{-1})^{ij}(p_j - \pi_j) - i\xi^i p_i/\hbar\bigr].\label{gauss1}
\end{eqnarray}
Also the formula (\ref{internal}) for the internal energy of the Gaussian state,
despite its appearance, exhibits a symmetry under the interchange of positions
and momenta. Denoting by $c$ and $d$ the real and the imaginary parts of the
matrix $K^{-1}$,
\begin{eqnarray}
c = (a + b a^{-1} b)^{-1},\;\;\;\;d = -a^{-1} b (a + b a^{-1} b)^{-1},
\end{eqnarray}
one can rewrite (\ref{energy}) as
\begin{eqnarray}
E_{\rm int} = \frac{\hbar}{4}{\rm Tr}\{u d c^{-1} d/m + u c/m + m g c^{-1}\}.
\label{energy1}
\end{eqnarray}

\section{Saturation of uncertainty relations\\ by squeezed states\label{ur}}

The special character of squeezed states comes forth clearly in the analysis of
uncertainty relations between positions and momenta. In order to simplify the
notation, I shall assume in this section that the expectation values
$\langle\hat x^i\rangle$ and $\langle\hat p_i\rangle$ are zero. In the simplest,
one-dimensional case the Schr\"odinger-Robertson uncertainty relation
\cite{schr,robert} has the form
\begin{eqnarray}
\Delta x^2 \Delta p^2 \geq \hbar^2/4 + \langle\hat x\hat p + \hat p\hat x\rangle^2/4.
\end{eqnarray}
This inequality is saturated by any Gaussian wave function. The equality
expressing the saturation of uncertainty relations in one dimension can also be
rewritten as a matrix equation
\begin{eqnarray}
Q J Q = (\hbar/4)^2 J,\label{simple}
\end{eqnarray}
where
\begin{eqnarray}
 Q = \left(
 \begin{array}{cc}
 \langle\hat x\hat x\rangle & \frac{1}{2}\langle\hat x\hat p + \hat p\hat
 x\rangle\\ \frac{1}{2}\langle\hat x\hat p + \hat p\hat x\rangle & \langle\hat
 p\hat p\rangle
 \end{array}\right),
\end{eqnarray}
and $J$ is the simplectic metric
\begin{eqnarray}
J =
 \left(
 \begin{array}{cc}
 0 & 1\\ -1 & 0
 \end{array}\right).
\end{eqnarray}
The set of conditions that correspond to the saturation of the
Schr\"odinger-Robertson inequality in the $N$-dimensional case has the same
general form (\ref{simple}) except that all matrix elements of the matrices $Q$
and $J$ become now $N\times N$ matrices
\begin{eqnarray}
Q_N = \left(
 \begin{array}{cc}
 \langle\hat x^i\hat x^j\rangle & \frac{1}{2}\langle\hat x^i\hat p_j + \hat
 p_j\hat x^i\rangle\\ \frac{1}{2}\langle\hat p_i\hat x^j + \hat x^j\hat
 p_i\rangle & \langle\hat p_i\hat p_j\rangle
 \end{array}\right),\label{qun}
\end{eqnarray}
and
\begin{eqnarray}
 J_N = \left(
 \begin{array}{cc}
 0 & \delta_j^{\,k}\\ -\delta^j_{\,k} & 0
 \end{array}\right).\label{jayn}
\end{eqnarray}
In order to prove the formula (\ref{simple}) in the $N$-dimensional case, one
may use the following expressions for the position and momenta correlations
functions that can be obtained from the formula (\ref{gauss}) by performing the
corresponding Gaussian integrals
\begin{eqnarray}
 \langle x^i x^j \rangle = \frac{\hbar}{2m}(a^{-1})^{ij},\label{corxx}\\ \langle
 p_i p_j \rangle = \frac{\hbar m}{2}(a + b a^{-1} b)_{ij},\label{corpp}\\
 \langle x^i p_j \rangle = \frac{i\hbar}{2}(\delta^i_{\,j} + i
 (a^{-1}b)^i_{\,j}), \label{corxp}\\ \langle p_i x^j \rangle =
 -\frac{i\hbar}{2}(\delta_i^{\,j} - i (ba^{-1})_i^{\,j}). \label{corpx}
\end{eqnarray}
These relations can easily be inverted to give $K$ and its inverse in terms of
correlations
\begin{eqnarray}
 K_{ik} = -i(\langle x^i x^j \rangle)^{-1}\langle x^jp_k\rangle/m,\\
 (K_{ik})^{-1} = i m(\langle p_i p_j \rangle)^{-1}\langle p_jx^k\rangle.
\end{eqnarray}
The validity of Eq. (\ref{simple}) with $Q$ and $J$ given by the expressions
(\ref{qun}) and (\ref{jayn}) can be checked now by a direct calculation with the
use of (\ref{corxx}--\ref{corpx}).

By a direct calculation one may also check that the entropic uncertainty
relation \cite{bbm}
\begin{eqnarray}
 -\int\!d^Nx\vert\psi({\bf r})\vert^2\ln\vert\psi({\bf r})\vert^2 -
 \int\!d^Np\vert\phi({\bf p})\vert^2\ln\vert\phi({\bf p})\vert^2 \ge N(\ln\pi +
 1)\label{bbm}
\end{eqnarray}
is saturated by all squeezed states. Note that the saturation of the
Schr\"odin\-ger-Robertson and the entropic uncertainty relations is a
characteristic feature that singles out all squeezed states. The whole set of
squeezed states can be characterized by these properties without any reference
to a specific Hamiltonian or to annihilation and creation operators.

\section{Time evolution of squeezed states\label{time_evol}}

The squeezed states retain their form during time evolution. In order to satisfy
the time-dependent Schr\"odinger equation, the vectors $\xi$, $\pi$, and the
matrix $K$ must be taken as functions of time, i.e.
\begin{eqnarray}
 \psi({\bf r},t) &=& ({\rm
 Det}a(t))^{1/4}(m/\pi\hbar)^{N/4}\exp(-i\xi^i\pi_i/2\hbar)\label{gauss_t}\\
 &\times&\exp\bigl[-\frac{m}{2\hbar}(x^i - \xi^i(t)K_{ij}(t))(x^j - \xi^j(t)) +
 i\pi_i(t) x^i/\hbar\bigr].\nonumber
\end{eqnarray}
Upon substituting this formula into the Schr\"odinger equation with the
Hamiltonian (\ref{hamiltonian}), one finds the classical equations of motion for
the position $\xi$ and momentum $\pi$ of a harmonic oscillator
\begin{eqnarray}
 \frac{d}{dt}\xi^i(t) = g^{ij}\pi_j(t),\;\;\;\;\; \frac{d}{dt}\pi_i(t) =
 -u_{ij}\xi^j(t),\label{evol_xp}
\end{eqnarray}
and the following nonlinear evolution equation for the matrix $K$
\begin{eqnarray}
 \frac{d}{dt}K(t) = -iK(t)\,m g\,K(t) + iu/m.\label{evol_k}
\end{eqnarray}
The vectors $\xi$ and $\pi$ determine the center of mass position and the center
of mass momentum of the Gaussian wave packet. The matrices $a$ and $b$ determine
the shape of the wave function and the distribution of the probability current
of the internal ``tumbling'' motion, respectively. It follows from Eq. (\ref{evol_k})
and from the relation $d K^{-1}/dt = -K^{-1} dK/dt K^{-1}$ that the inverse matrix
$K^{-1}$ satisfies the evolution equation in which the roles of $g$ and $u$ are
interchanged,
\begin{eqnarray}
 \frac{d}{dt}K^{-1}(t) = -iK^{-1}(t)\,u/m\,K^{-1}(t) + i m g,\label{evol_k1}
\end{eqnarray}
as one would expect from the formula (\ref{gauss1}) and the symmetry of the
Hamiltonian under the interchange of positions and momenta.

The nonlinear equation (\ref{evol_k}) is a matrix Riccatti equation and it may
be converted into a set of linear equations by the following substitution ($g =
1/m$)
\begin{eqnarray}
 K(t) = D^{-1}(t)N(t)\Omega.\label{subst}
\end{eqnarray}
Eq. (\ref{evol_k}) is satisfied when two complex matrices $N$ and $D$ satisfy
the set of equations
\begin{eqnarray}
 \frac{d}{dt}D(t) = iN(t)\Omega,\;\;\;\; \frac{d}{dt}N(t) =
 iD(t)\Omega.\label{evol_nd}
\end{eqnarray}
The solution of this linear set of equations
\begin{eqnarray}
 N(t) &=& N(0)\cos\Omega t + iD(0)\sin\Omega t,\label{n_t}\\ D(t) &=&
 D(0)\cos\Omega t + iN(0)\sin\Omega t,\label{d_t}
\end{eqnarray}
gives the following result for $K(t)$
\begin{eqnarray}
 K(t) = (\cos\Omega t + iK(0) \frac{\sin\Omega t}{\Omega})^{-1} (K(0) \cos\Omega
 t + i\Omega\sin\Omega t).\label{k_t}
\end{eqnarray}
The order of matrices is here important since in general the initial value
$K(0)$ and $\Omega$ do not commute. Actually, there are two equivalent forms of
$K(t)$ resulting from the identity
\begin{eqnarray}
&&(\cos\Omega t + iK(0) \frac{\sin\Omega t}{\Omega})^{-1}
 (K(0) \cos\Omega t + i\Omega\sin\Omega t)\nonumber\\ &=& (\cos\Omega t\;K(0) +
 i\Omega\sin\Omega t) (\cos\Omega t + i\frac{\sin\Omega
 t}{\Omega}K(0))^{-1}.\label{equality}
\end{eqnarray}
Since both sides of this equality are related by the matrix transposition, the
symmetry of the matrix $K(t)$ is preserved during time evolution. For the ground
state (and only for the ground state), $K$ is constant in time, even though both
matrices $N$ and $D$ vary with time: $N(t) = D(t) = e^{i\Omega t}$. From Eq.
(\ref{k_t}) one obtains the following formula for the time evolution of $a$
\begin{eqnarray}
 a(t) = D^{-1}(t)a_0D^{\dagger-1}(t),\label{a_t}
\end{eqnarray}
that shows explicitly that the positive definite character of $a$ is preserved
during the time evolution.

The formula (\ref{k_t}) describes completely the dynamics of squeezed states for
an arbitrary harmonic oscillator. The time evolution of $K$ may get quite
complicated for a multidimensional harmonic oscillator when the matrices
$\Omega$ and $K(0)$ do not commute so that they cannot be simultaneously
diagonalized. In the simple case, when the initial value of $K$ represents just
a small perturbation of the ground state,
\begin{eqnarray}
K(0) = \Omega + \delta K(0),
\end{eqnarray}
the formula (\ref{k_t}) simplifies as follows
\begin{eqnarray}
 K(t) = \Omega + e^{-i\Omega t} \,\delta K(0) e^{-i\Omega t}.
 \label{first_order}
\end{eqnarray}
An alternative method of finding the time evolution will be given in the next
section in terms of the Wigner function.

\section{Wigner function of squeezed states\label{wigner}}

The Wigner function $W_{\rm sq}({\bf r}, {\bf p})$ of a general squeezed state
(\ref{gauss}) can be easily determined from the defining formula \cite{wigner}
\begin{eqnarray}
W({\bf r}, {\bf p}) = 2^{-N}\int\!d^N\eta \exp(i \mbox{\boldmath $\eta$}
 \!\cdot\!{\bf p}/\hbar) \psi({\bf r} - \mbox{\boldmath $\eta$}/2) \psi^*({\bf
 r} + \mbox{\boldmath $\eta$}/2)\label{wig}
\end{eqnarray}
by evaluating an $N$-dimensional Gaussian integral. The prefactor in (\ref{wig})
has been chosen so that the Wigner function is a pure exponential,
\begin{eqnarray}
W_{\rm sq}({\bf r}, {\bf p}) = \exp\lbrack-\frac{1}{\hbar}(x - \xi)^i
 (a + b a^{-1} b)_{ij}(x - \xi)^j\rbrack\nonumber\\ \times\exp\lbrack -
 \frac{1}{\hbar}(p - \pi)_i (a^{-1})^{ij}(p - \pi)_j - \frac{2}{\hbar}(x -
 \xi)^i (b a^{-1})_i^{\;j} (p - \pi)_j\rbrack.
\end{eqnarray}
This formula can be rewritten in a compact form in terms of the expectation
values of the position and momentum operators
\begin{eqnarray}
 W_{\rm sq}({\bf r}, {\bf p}) = \exp\lbrack-\frac{2}{\hbar^2}
 \langle\left((x^i-\langle\hat x^i\rangle)\hat p_i - (p_i-\langle\hat
 p_i\rangle)\hat x^i\right)^2\rangle\rbrack.\label{wig_exp}
\end{eqnarray}
Thus, the expectation values of the position and momenta and their quadratic
forms determine the Wigner function. This was to be expected since every
Gaussian distribution is determined by its second moments. For the squeezed
states centered at the origin, this formula takes on the form
\begin{eqnarray}
 W_{\rm sq}({\bf r}, {\bf p}) = \exp\lbrack-\frac{2}{\hbar^2} \lbrack
 x^i\langle\hat p_i\hat p_j\rangle x^j + p_i\langle\hat x^i\hat x^j\rangle p_j -
 p_i\langle\hat x^ip_j + \hat p_j\hat x^i\rangle x^j\rbrack\rbrack.\label{rep}
\end{eqnarray}
The time evolution of the Wigner function of a harmonic oscillator is determined
by the equation
\begin{eqnarray}
 \partial_t W({\bf r}, {\bf p},t) = -(p_i g^{ij}\nabla_j - x^i
 u_{ij}\partial^j)W({\bf r}, {\bf p},t),\label{time_wig}
\end{eqnarray}
where $\nabla$ and $\partial$ denote the differentiation with respect to ${\bf
r}$ and ${\bf p}$, respectively. The solution of this equation can be written in
the form
\begin{eqnarray}
 W({\bf r}, {\bf p},t) = W({\bf r}(-t), {\bf p}(-t),t=0),
\end{eqnarray}
where $({\bf r}(-t), {\bf p}(-t))$ is the time-reversed solution of the
classical equations of motion. Owing to the invariance of the simplectic form
$x^i \bar p_i - p_i\bar x^i$ under all canonical transformations including the
time translation, the formula (\ref{rep}) for the time-dependent Wigner function
of a squeezed state can also be written in terms of the solutions of the
Heisenberg equations of motion for the position and momentum operators
\begin{eqnarray}
 W_{\rm sc }({\bf r}, {\bf p},t) = \exp\lbrack-\frac{2}{\hbar^2}
 \langle\left((x^i-\langle\hat x^i(t)\rangle)\hat p_i(t) - (p_i-\langle\hat
 p_i(t)\rangle)\hat x^i(t)\right)^2\rangle\rbrack.
\end{eqnarray}

\section{Quantized electromagnetic field\label{qed}}

Owing to the correspondence between a set of harmonic oscillators and the
electromagnetic field, an analogous approach to the one presented here works, at
least at the formal level, for the general squeezed states of the
electromagnetic field.

In order to use the analogy with quantum mechanics of harmonic oscillators to
its full extent, I shall employ the formulation of quantum electrodynamics in a
representation which is the closest possible counterpart of the Schr\"odinger
representation. Jackiw, who is also fond of this approach, called it in Ref.
\cite{jackiw} the {\em field theoretic Schr\"odinger representation}. In this
representation the state is described by a functional $\Psi\left[{\bf A}\right]$
of the vector potential. The vector potential operator $\hat{\bf A}({\bf r})$
and the magnetic induction operator $\hat{\bf B}({\bf r})$ acts on the state
functional as a multiplication,
\begin{eqnarray}
 \hat{\bf A}({\bf r})\Psi\left[{\bf A}\right] = {\bf A}({\bf r})\Psi\left[{\bf
 A}\right],\;\;\;\; \hat{\bf B}({\bf r})\Psi\left[{\bf A}\right] =
 \bigl(\nabla\times{\bf A}({\bf r})\bigr)\Psi\left[{\bf
 A}\right],\label{action_ab}
\end{eqnarray}
while the electric displacement operator $\hat{\bf D}({\bf r})$ acts as the
functional differentiation,
\begin{eqnarray}
 \hat{\bf D}({\bf r})\Psi\left[{\bf A}\right] = i\hbar\frac{\delta}{\delta{\bf
 A}({\bf r})} \Psi\left[{\bf A}\right]. \label{action_d}
\end{eqnarray}
The Hamiltonian in this representation takes on the form
\begin{eqnarray}
 H = \frac{1}{2}\int\!\!d^3r\, \bigl[-\frac{\hbar^2}{\epsilon({\bf
 r})}\frac{\delta^2}{\delta{\bf A} ({\bf r})^2} + \frac{1}{\mu({\bf
 r})}\bigl(\nabla\times{\bf A}({\bf r})\bigr)^2\bigr]. \label{ham_em}
\end{eqnarray}
This representation is the analog of the Schr\"odinger (position)
representation. There is also the analog of the momentum representation in which
the electric displacement vector acts as a multiplication and the vector
potential acts as the functional derivative.

The most general Gaussian functional is determined by two vector functions
${\cal A}({\bf r})$ and ${\cal D}({\bf r})$ and --- the counterparts of $\xi$
and $\pi$ --- and a complex symmetric kernel ${\cal K}({\bf r},{\bf r}')$ ---
the counterpart of the matrix $K$,
\begin{eqnarray}
 \Psi\left[{\bf A}\right] &=&
 C\exp\bigl[-\frac{1}{2\hbar}\int\!\!d^3r\!\!\int\!\!d^3r'\, ({\bf A}({\bf r}) -
 {\cal A}({\bf r}))\!\cdot\!{\cal K}({\bf r},{\bf r}') \!\cdot\!({\bf A}({\bf
 r}') - {\cal A}({\bf r}'))\bigr]\nonumber\\
 &&\times\exp\bigl[\frac{i}{\hbar}\int\!\!d^3r\,{\cal D}({\bf r})\!\cdot\! {\bf
 A}({\bf r})\bigr], \label{gaussian_a}
\end{eqnarray}
In addition to being a function of two vector arguments, the kernel ${\cal K}$
is also a $3\times 3$ matrix since it acts on the vector indices of $A_i$.
The vector function ${\cal A}({\bf r})$ and the kernel ${\cal K}$ are just
labels characterizing the state. To distinguish ${\cal A}$ from the argument
${\bf A}$ of the state functional, they are set in a different font. In order to
secure gauge invariance, i.e., the invariance of $\Psi\left[{\bf A}\right]$
under the gauge transformations ${\bf A}({\bf r}) \to {\bf A}({\bf r}) +
\nabla\lambda({\bf r})$, the kernel ${\cal K}^{ij}({\bf r},{\bf r}')$ must obey
the conditions
\begin{eqnarray}
 \partial_i{\cal K}^{ij}({\bf r},{\bf r}') = 0 = {\cal K}^{ij}({\bf r},{\bf
 r}')\lpt_j'.\label{condition}
\end{eqnarray}
Thus, the kernel ${\cal K}$ must be a double curl of some new kernel ${\cal W}$,
\begin{eqnarray}
 {\cal K}^{ij}({\bf r},{\bf r}') = \epsilon^{ikm}\partial_k{\cal W}_{mn}({\bf
 r},{\bf r}')\lpt_l'\epsilon^{jln}. \label{new_kernel}
\end{eqnarray}
This means that $\Psi\left[{\bf A}\right]$ depends on ${\bf A}({\bf r})$ only
through ${\bf B}({\bf r}) = \nabla\times{\bf A}({\bf r})$.

The total energy of the electromagnetic field, like in the case of the
quantum-mechanical oscillator, is made of two parts: the classical energy of the
coherent field and the quantum correction due to squeezing
\begin{eqnarray}
 E &=& \frac{1}{2}\int\!\!d^3r\left(\frac{{\cal D}^2({\bf r})}{\epsilon({\bf
 r})}\ + \frac{{\cal B}^2({\bf r})}{\mu({\bf r})}\right)\\ &+&
 \frac{\hbar}{4}\int\!\!d^3r\!\!\int\!\!d^3r' \frac{\delta_{kl}\delta({\bf r} -
 {\bf r}')} {\mu({\bf r})}\epsilon^{imk}\partial_m a^{-1}_{ij}({\bf r},{\bf r}')
 \epsilon^{jnl}\lpt_n'\nonumber\\ &+& \frac{\hbar}{4}\int\!\!d^3r\!\!\
 \frac{\delta_{ij}}{\epsilon({\bf r})}\left(\int\!\!d^3r'\int\!\!d^3r''
 b^{ik}({\bf r},{\bf r}')a^{-1}_{kl}({\bf r}',{\bf r}'') b^{lj}({\bf r}'',{\bf
 r}) + a^{ij} ({\bf r},{\bf r})\right),\nonumber
\end{eqnarray}
where $a^{ij}$ and $b^{ij}$ are the real and the imaginary part of ${\cal
K}^{ij}$,
\begin{eqnarray}
{\cal K}^{ij}({\bf r},{\bf r}') = a^{ij}({\bf r},{\bf r}')
 + i b^{ij}({\bf r},{\bf r}')
\end{eqnarray}
and the inverse of $a^{ij}$ is to be taken in the subspace of transverse kernels
satisfying the conditions (\ref{condition}).

The analog of the equation (\ref{ground_eq}) that determines the ground state of
the harmonic oscillator in quantum mechanics has now the form
\begin{eqnarray}
 \int\!\!d^3r\, {\cal K}^{ik}({\bf r}',{\bf
 r},t)\frac{\delta_{kl}}{\epsilon({\bf r})} {\cal K}^{lj}({\bf r},{\bf r}'',t) =
 \epsilon^{imk}\partial_m'\frac{\delta_{kl} \delta({\bf r}' - {\bf
 r}'')}{\mu({\bf r}')}\lpt_n''\epsilon^{jnl}. \label{ground_eq1}
\end{eqnarray}

In the simplest case, when the medium is homogeneous, the kernel ${\cal K}$ may
depend only on the difference of coordinates and the solution of Eq.
(\ref{ground_eq1}) can be easily found by the Fourier transformation. The
resulting expression has the form
\begin{eqnarray}
 \tilde{\cal K}^{ij}({\bf k}) =
 \sqrt{\frac{\epsilon}{\mu}}\epsilon^{imk}k_m\frac{\delta_{kl}} {\vert{\bf
 k}\vert}k_n\epsilon^{jnl}.\label{free_k}
\end{eqnarray}
The coordinate representation of this expression is
\begin{eqnarray}
 {\cal K}^{ij}({\bf r} - {\bf r}') = \frac{1}{4\pi^2}
 \sqrt{\frac{\epsilon}{\mu}}\epsilon^{ikm}\partial_k
 \frac{\delta_{mn}}{\vert{\bf r} - {\bf r}'\vert^2}\lpt_l'\epsilon^{jln}.
 \label{vac_kernel}
\end{eqnarray}
Therefore, the ground-state functional of the electromagnetic field in the
homogeneous medium is
\begin{eqnarray}
 \Psi_0\left[{\bf A}\right] = C\exp\left[-\frac{1}{4\pi^2\hbar}
 \sqrt{\frac{\epsilon}{\mu}}\int\!\!d^3r\!\!\int\!\!d^3r'\, {\bf B}({\bf
 r})\frac{1}{\vert{\bf r} - {\bf r}'\vert^2} {\bf B}({\bf
 r}')\right].\label{ground_m}
\end{eqnarray}
This formula in the case of the vacuum state has been written down by Wheeler
\cite{wheeler}. Here I would like to study its significance in the presence of a
medium. The existence of the ground state in any static medium enables one to
define the notion of squeezing with respect to this ground state. One can see
from (\ref{ground_m}) that the dielectrics decrease the fluctuations of the
magnetic field, while the magnetics increase them.

In the case of the electromagnetic field there also exists an analog of the
quantum-mechanical momentum representation. In this representation the state of
the electromagnetic field is given as a functional of ${\bf D}$ which is the
canonically conjugate variable to the potential ${\bf A}$. Owing to the known
symmetry of Maxwell theory without sources, the formula for the ground state
functional in the ${\bf D}$ representation can be obtained by the replacements
\begin{eqnarray}
{\bf D}\;\to\;{\bf B},\;\;{\bf B}\;\to\;-{\bf
D},\;\;\epsilon\;\to\;\mu,\;\;\mu\;\to\;\epsilon.
\end{eqnarray}
and it has the form
\begin{eqnarray}
 \Psi_0\left[\tilde{\bf A}\right] = C'\exp\left[-\frac{1}{4\pi^2\hbar}
 \sqrt{\frac{\mu}{\epsilon}}\int\!\!d^3r\!\!\int\!\!d^3r'\, {\bf D}({\bf
 r})\frac{1}{\vert{\bf r} - {\bf r}'\vert^2} {\bf D}({\bf
 r}')\right],\label{ground_e}
\end{eqnarray}
where $\tilde{\bf A}$ is the vector potential for ${\bf D}$, i.e. ${\bf D} =
\nabla\times\tilde{\bf A}$.

For an inhomogeneous medium, one may easily find first-order changes in ${\cal
K}$ by perturbation theory. The most interesting case is that of a change in the
dielectric constant. The exact kernel for the ground state in a static medium
satisfies the equation
\begin{eqnarray}
 \int\!\!d^3r\,{\cal K}^{ik}({\bf r}',{\bf r})\frac{\delta_{kl}}{\epsilon({\bf
 r})} {\cal K}^{lj}({\bf r},{\bf r}'')
  = \epsilon^{imk}\partial_m'\frac{\delta_{kl} \delta({\bf r}' - {\bf
 r}'')}{\mu({\bf r}')}\lpt_n''\epsilon^{jnl}, \label{vacuum_k}
\end{eqnarray}
from which the following equation is obtained by taking a variation due a small
change in $\epsilon({\bf r})$
\begin{eqnarray}
 &&\int\!\!d^3r\,\delta{\cal K}^{ik}({\bf r}',{\bf r})
 \frac{\delta_{kl}}{\epsilon({\bf r})} {\cal K}^{lj}({\bf r},{\bf r}'') +
 \int\!\!d^3r\,{\cal K}^{ik}({\bf r}',{\bf r}) \frac{\delta_{kl}}{\epsilon({\bf
 r})} \delta{\cal K}^{lj}({\bf r},{\bf r}'') \nonumber\\ &=& \int\!\!d^3r\,{\cal
 K}^{ik}({\bf r}',{\bf r}) \frac{\delta_{kl}\delta{\epsilon}({\bf r})}
 {\epsilon({\bf r})^2}{\cal K}^{lj}({\bf r},{\bf r}''). \label{var_eps}
\end{eqnarray}
When the variation is taken around the vacuum state, the translational
invariance of this state allows one to seek the change in ${\cal K}$ in the form
\begin{eqnarray}
 \delta{\cal K}^{ij}({\bf r}',{\bf r}'') = \int\!\!d^3r\,\Gamma^{ij}({\bf r}' -
 {\bf r},{\bf r} - {\bf r}'')\delta\epsilon({\bf r}). \label{delta_k}
 \label{gamma_r}
\end{eqnarray}
Applying the Fourier transformation to Eq. (\ref{var_eps}), one finds the
following Fourier transform of the kernel $\Gamma^{ij}$
\begin{eqnarray}
 {\tilde\Gamma}^{ij}({\bf k}_1,{\bf k}_2) = (\delta^{im}{\bf k}^2_1 - k^i_1
k^m_1) \frac{c\delta_{mn}}{\vert{\bf k}_1 \vert \vert{\bf k}_2\vert (\vert{\bf
k}_1\vert + \vert{\bf k}_2\vert)} (\delta^{nj}{\bf k}_2^2 - k^n_2
k^j_2),\label{fourier_g}
\end{eqnarray}
where $c$ is the speed of light in the medium. In the coordinate representation,
the expression for $\Gamma^{ij}$ reads
\begin{eqnarray}
 \Gamma^{ij}({\bf r}_1,{\bf r}_2) = (\delta^{ik}\Delta - \partial^i\partial^k)_1
 \frac{c\delta_{kl}}{(2\pi)^3\vert{\bf r}_1\vert\vert{\bf r}_2\vert (\vert{\bf
 r}_1\vert + \vert{\bf r}_2\vert)} (\delta^{lj}\llapl -
 \lpt^l\lpt^j)_2.\label{gamma}
\end{eqnarray}
This expression shows the long range effect in the squeezing produced by a
dielectric: when one moves away from the dielectric, the function $\Gamma$ falls
only as $1/r^4$.

With the use of Feynman-Hellmann theorem (\ref{hell_feyn}) one may directly
arrive at the first order correction to the ground state energy $E_0$ that is
due to a small departure of the dielectric constant from its vacuum value
$\epsilon_0$, without ever calculating $\delta K$. To this end, I shall use the
following adaptation to the present case of the formula (\ref{hell_feyn}) for
the harmonic oscillator
\begin{eqnarray}
 \delta E_0 &=& -\frac{\hbar}{4}\int\!\!d^3r\!\!\int\!\!d^3r'
 \frac{\delta\epsilon({\bf r})}{\epsilon^2_0}\delta_{ij} K^{ij}({\bf r}' - {\bf
 r},{\bf r} - {\bf r}') \nonumber\\ &=&
 -\frac{1}{4}\int\!\!d^3r\,\frac{\delta\epsilon({\bf r})}{\epsilon_0}
 2\!\int\!\!\frac{d^3k}{(2\pi)^3}\hbar c\vert{\bf k}\vert.\label{delta_e}
\end{eqnarray}
The integral over ${\bf k}$ represents the energy of all photons that are affected
by the change in the dielectric constant and the factor of 2 comes from two
polarization states of the photon. Obviously, the integration does not extend to
infinity because all dielectrics become transparent (behave like the vacuum) for
sufficiently high photon frequency and that fact will introduce a cutoff.
Therefore, a more appropriate way to write Eq. (\ref{delta_e}) is
\begin{eqnarray}
 \delta E_0 = -\frac{1}{4}\int\!\!d^3r\,2\!\int\!\!\frac{d^3k}{(2\pi)^3}\,
 \frac{\delta\epsilon_{\bf k}({\bf r})}{\epsilon_0} \hbar c_{\bf k}({\bf
 r})\vert{\bf k}\vert, \label{delta_e1}
\end{eqnarray}
where $\epsilon$ and $c$ are taken as slowly varying functions of the wave
vector ${\bf k}$. Since $\delta c= \delta (1/\sqrt{\epsilon\mu}) = -
(\delta\epsilon/2\epsilon) c$, the expression (\ref{delta_e1}) can be recognized
as the variation of the zero point energy evaluated in the local approximation
when the variation of $\epsilon$ in space is very slow over the distance of a
wave length $1/k$,
\begin{eqnarray}
 \delta E_0 =
 \delta\bigl(\frac{1}{2}\int\!\!d^3r\,2\!\int\!\!\frac{d^3k}{(2\pi)^3}\, \hbar
 c_{\bf k}({\bf r})\vert{\bf k}\vert\bigr).
\end{eqnarray}

First order perturbation theory, described here as a simple illustration of the
global approach to squeezing, is clearly not sufficient to give the Casimir
force. This force can be understood as coming from the change in the energy of
the ground state but it requires the calculation of the energy at least up to
second order. In the first order, one obtains only the volume effect. In
particular, the change of the relative positions of several pieces of the
dielectric would not lead to a change in the energy and therefore would not give
rise to an interaction force.

\section{Propagation of squeezing\label{prop}}

In order to satisfy the time-dependent Schr\"odinger equation
\begin{eqnarray}
 i\hbar\partial_t\Psi\left[{\bf A}\vert t\right] = H\Psi\left[{\bf A}\vert
 t\right],\label{schrod}
\end{eqnarray}
the functions ${\cal A}$ and ${\cal D}$ and the kernel ${\cal K}$ must depend on
time, except for the ground state. Their time evolution can be derived by
substituting the wave functional
\begin{eqnarray}
 &&\Psi\left[{\bf A}\vert t\right]\nonumber\\ && =
 C\exp\bigl[-\frac{1}{2\hbar}\int\!\!d^3r\int\!\!d^3r'\, ({\bf A}({\bf r}) -
 {\cal A}({\bf r},t))\!\cdot\!{\cal K}({\bf r},{\bf r}',t) \!\cdot\!({\bf
 A}({\bf r}') - {\cal A}({\bf r}',t))\bigr]\nonumber\\
 &&+\exp\bigl[\frac{i}{\hbar}\int\!\!d^3r\,{\cal D}({\bf r},t)\!\cdot\! {\bf
 A}({\bf r})\bigr]\label{gaussian_at}
\end{eqnarray}
into (\ref{schrod}) and then by comparing the terms of the same form on both
sides of this equation. In this way one obtains the Maxwell equations for ${\cal
B} = \nabla\times{\cal A}$ and ${\cal D}$
\begin{eqnarray}
 \partial_t{\cal B}({\bf r},t) = -\nabla\times\frac{{\cal D}({\bf
 r},t)}{\epsilon({\bf r})},\\ \partial_t{\cal D}({\bf r},t) =
 \nabla\times\frac{{\cal B}({\bf r},t)}{\mu({\bf r})},
\end{eqnarray}
and the following nonlinear integro-differential equation for ${\cal K}$
\begin{eqnarray}
 &&\partial_t{\cal K}^{ij}({\bf r}',{\bf r}'',t) \\ &&= -i\int\!\!d^3r\, {\cal
 K}^{ik}({\bf r}',{\bf r},t)\frac{\delta_{kl}}{\epsilon({\bf r})} {\cal
 K}^{lj}({\bf r},{\bf r}'',t) + i\epsilon^{imk}\partial_m'\frac{\delta_{kl}
 \delta({\bf r}' - {\bf r}'')}{\mu({\bf r}')}\lpt_n''\epsilon^{jnl}.\nonumber
 \label{evolution_k}
\end{eqnarray}

Propagation of squeezing, in principle, is fully described by the time evolution
equation (\ref{k_t}) of the kernel ${\cal K}$ but it is rather difficult to draw
physical conclusions from this equation in the general case. However, for small
perturbations, when the departure from the vacuum situation is given by
(\ref{delta_k}) and (\ref{gamma}), one can find an explicit expression for the
time dependence of ${\cal K}$. To this end, I shall use the first order formula
(\ref{first_order}) obtained for the harmonic oscillator. In the present case,
this formula takes on a very simple form when ${\cal K}$ is taken as a function of
wave vectors because in this representation the frequency $\Omega$ is diagonal.
The application of the operations $\exp(-i\Omega t)$ in the formula
(\ref{first_order}) amounts simply to the multiplications by $\exp(-i c\vert{\bf
k}\vert t)$. This leads to the following formula for $\delta{\cal K}(t)$
\begin{eqnarray}
&& \delta{\cal K}^{ij}({\bf r}_1,{\bf r}_2, t)\nonumber\\
 &&= \int\!\!d^3r\int\!\!\frac{d^3k_1}{(2\pi)^3}\frac{d^3k_2}{(2\pi)^3} e^{i
 {\bf k}_1\!\cdot\!({\bf r}_1 - {\bf r})} e^{i {\bf k}_2\!\cdot\!({\bf r} - {\bf
 r}_2)} \delta\epsilon({\bf r})e^{-i c\vert{\bf k}_1\vert t} e^{-i c \vert{\bf
 k}_2\vert t} \tilde\Gamma^{ij}({\bf k}_1,{\bf k}_2)\nonumber\\ &&= \int\!\!d^3r
 \Gamma^{ij}({\bf r}_1 - {\bf r}, {\bf r} - {\bf r}_2, t) \delta\epsilon({\bf
 r}),
\end{eqnarray}
The function $\Gamma({\bf r}_1 - {\bf r}, {\bf r} - {\bf r}_2, t)$ can be easily
evaluated explicitly since after the substitution of the expression
(\ref{fourier_g}) all integrations over the wave vectors are elementary. The
resulting formula reads
\begin{eqnarray}
 &&\Gamma^{ij}({\bf r}_1 - {\bf r}, {\bf r} - {\bf r}_2, t) \nonumber\\ &&= c
 (\delta^{ik}\Delta - \partial^i\partial^k)_1 \delta_{kl}G({\bf r}_1 - {\bf r},
 {\bf r} - {\bf r}_2, t)(\delta^{lj}\llapl - \lpt^l\lpt^j)_2,
\end{eqnarray}
where
\begin{eqnarray}
G({\bf r}_1, {\bf r}_2, t) &=&
\int\!\!\frac{d^3k_1}{(2\pi)^3}\frac{d^3k_2}{(2\pi)^3}
 e^{i {\bf k}_1\!\cdot\!{\bf r}_1} e^{i {\bf k}_2\cdot{\bf r}_2}
\frac{ e^{-i c\vert{\bf k}_1\vert t} e^{-i c \vert{\bf k}_2\vert t}}{\vert{\bf k}_1 \vert \vert{\bf k}_2\vert
 (\vert{\bf k}_1\vert + \vert{\bf k}_2\vert)} \\ &=& \frac{1}{(2
\pi^2)^2}\int_0^\infty\!\!dk_1\int_0^\infty\!\!dk_2 \frac{ e^{-i c k_1 t} e^{-i
c k_2 t}} {k_1 + k_2} \frac{\sin(k_1 r_1)}{r_1} \frac{\sin(k_2
r_2)}{r_2}.\nonumber
\end{eqnarray}
The integration over $k_1$ and $k_2$ can be easily performed after the change of
variables: $k = k_1 + k_2, \; q = k_1 - k_2$
\begin{eqnarray}
&&G({\bf r}_1, {\bf r}_2, t) =
 \frac{1}{(2 \pi)^3 \pi r_1 r_2}\int_0^\infty\frac{dk}{k} e^{-i k c t}
 \int_{-k}^k\!\!dq \sin\frac{k+q}{2}r_1 \sin\frac{k-q}{2}r_2\nonumber\\ &&=
 \frac{2}{(2 \pi)^3 \pi (r_1^2 - r_2^2)}\int_0^\infty\frac{dk}{k} e^{-i k c t}
 \left(\frac{\sin kr_2}{r_2} - \frac{\sin kr_1}{r_1}\right)\label{propagation}\\
 &&= \frac{1}{(2 \pi)^3(r_1^2 - r_2^2)} \nonumber\\
 &&\times\left[\frac{\theta(r_2 - \vert c t\vert)}{r_2} - \frac{\theta(r_1 -
 \vert c t\vert)}{r_1} + \frac{i}{\pi r_2}\ln\left\vert\frac{r_2 - c t}{r_2 + c
 t}\right\vert - \frac{i}{\pi r_1}\ln\left\vert\frac{r_1 - c t}{r_1 + c
 t}\right\vert\right],\nonumber
\end{eqnarray}
where $\theta(r - ct)$ is the step function. The function (\ref{propagation})
describes fully the propagation of squeezing in the vacuum when the departure
from the vacuum state is small (cf. Eq. (\ref{first_order})). Unfortunately,
this approximation is not valid near the light cone in the variables $r$ and $t$
because of the singularities in Eq. (\ref{propagation}). A better,
nonperturbative approach is needed to determine the behavior of squeezing on the
light cone. What does follow, however, from these considerations is that
squeezing has very interesting propagation properties. In my simple example, the
initial value of the squeezing kernel ${\cal K}$ has been influenced by a change
in the value of $\epsilon$ that took place in some region of space. This
influence extends throughout space with the power-law falloff when moving away
from the region where $\epsilon$ has been changed, as determined by the formula
(\ref{gamma}). Then, at $t = 0$, the vacuum value of  $\epsilon$ is restored and
the electromagnetic field begins its relaxation to the true vacuum state. The
real part of ${\cal K}$ relaxes to zero with the speed of light as seen from the
presence of the step functions, while the imaginary part relaxes to zero much
more slowly as determined by the logarithmic functions.

\section{Wigner functional for the electromagnetic\\ field and relativistic invariance\label{wig_rel}}

There is a tendency among some physicists to limit the notion of
relativistic invariance to theories that involve only tensorial
(or spinorial) equations. The case of the Wigner functional and
the number of photons, that is explored in this section, shows
clearly that this limitation is not justified and that it
restricts unnecessarily the freedom of choosing the right
formulation. All that is really required of a relativistically
invariant theory is that the physical laws are the same in all
inertial reference frames. This implies that the equations used by
all observers related by a Poincar\'e transformation have the same
form. The exact nature of the transformation laws for physically
relevant quantities is of no importance if they only guarantee the
correct final result. There is no simple way to write the
expression for the Wigner functional for the vacuum state in terms
of tensor quantities with properly summed up indices even though
this functional is an absolute invariant. It has the same value
and the same form for every observer and it is even invariant
\cite{gross} under conformal transformations.

The description of squeezed states of the electromagnetic field in terms of the
Wigner function can be obtained by analogy with quantum mechanics. As a starting
point of this generalization I shall take the formula (\ref{wig_exp}). The
generalization consists of replacing the position ${\bf r}$ by the vector potential and
the momentum ${\bf p}$ by (minus) the electric displacement vector. In this way, one
obtains the following functional of ${\bf A}$ and ${\bf D}$ describing the most
general squeezed state of the electromagnetic field
\begin{eqnarray}
&&W_{\rm sq}\lbrack{\bf A},{\bf D}\rbrack\\
&&=\exp\bigl[-\frac{2}{\hbar^2}\!\int\!\!d^3r\!\!\int\!\!d^3r'\,
 ({\bf A}({\bf r}) - \langle{\bf\hat A}({\bf r})\rangle)
\!\cdot\!\langle{\bf\hat D}({\bf r}){\bf\hat D}({\bf r}')\rangle \!\cdot\!({\bf
A}({\bf r}') - \langle{\bf\hat A}({\bf r}')\rangle)\bigr]\nonumber\\
&&\times\exp\bigl[-\frac{2}{\hbar^2}\!\int\!\!d^3r\!\!\int\!\!d^3r'\,
 ({\bf D}({\bf r}) - \langle{\bf\hat D}({\bf r})\rangle)
\!\cdot\!\langle{\bf\hat A}({\bf r}){\bf\hat A}({\bf r}')\rangle \!\cdot\!({\bf
D}({\bf r}') - \langle{\bf\hat D}({\bf r}')\rangle)\bigr]\nonumber\\
&&\times\exp\bigl[-\frac{2}{\hbar^2}\!\int\!\!d^3r\!\!\int\!\!d^3r'\,
 ({\bf A}({\bf r}) - \langle{\bf\hat A}({\bf r})\rangle)
\!\cdot\!\langle{\bf\hat D}({\bf r}){\bf\hat A}({\bf r}')\rangle \!\cdot\!({\bf
D}({\bf r}') - \langle{\bf\hat D}({\bf r}')\rangle)\bigr]\nonumber\\
&&\times\exp\bigl[-\frac{2}{\hbar^2}\!\int\!\!d^3r\!\!\int\!\!d^3r'\,
 ({\bf D}({\bf r}) - \langle{\bf\hat D}({\bf r})\rangle)
 \!\cdot\!\langle{\bf\hat A}({\bf r}){\bf\hat D}({\bf r}')\rangle \!\cdot\!({\bf
 A}({\bf r}') - \langle{\bf\hat A}({\bf r}')\rangle)\bigr].\nonumber
 \label{wig_em}
\end{eqnarray}
Owing to gauge invariance, the Wigner functional depends on the
vector potential ${\bf A}$ only through the magnetic induction
vector ${\bf B}$ and I shall take this explicitly into account by
treating $W$ as a functional of ${\bf B}$ and ${\bf D}$. It should
be stressed that the Wigner functional for the electromagnetic
field is quite different from the Wigner function for the photon
wave function discussed in Refs. \cite{ibb4,ibb5}. The Wigner
functional describes the quantum statistical properties of the
full electromagnetic field and not just of a single photon.

In the simplest case of the vacuum state of the electromagnetic field the
expectation values of field operators can easily be evaluated and the resulting
formula for the Wigner functional can be written in a form resembling the
functionals representing the vacuum state vectors (\ref{ground_m}) and
(\ref{ground_e})
\begin{eqnarray}
 W_{\rm vac }\left[{\bf B},{\bf D}\right] = \exp(-2 N\left[{\bf B},{\bf
 D}\right]),
\end{eqnarray}
where
\begin{eqnarray}
&&\!\!N\left[{\bf B},{\bf D}\right]\label{phot_n}\\ &&\!\!=
\frac{1}{4\pi^2\hbar}
 \int\!\!d^3r\!\!\int\!\!d^3r'\, \left(\sqrt{\frac{\epsilon}{\mu}}{\bf B}({\bf
 r}) \frac{1}{\vert{\bf r}-{\bf r}'\vert^2} {\bf B}({\bf r}') +
 \sqrt{\frac{\mu}{\epsilon}}{\bf D}({\bf r}) \frac{1}{\vert{\bf r}-{\bf
 r}'\vert^2}{\bf D}({\bf r}')\right).\nonumber
\end{eqnarray}
The physical significance of the functional $ N\left[{\bf B},{\bf D}\right]$ has
been discovered by Zeldovich \cite{zeld} long time ago. It is the average number
of photons present in the electromagnetic field described by the vectors ${\bf
B}$ and ${\bf D}$. This interpretation can be justified by noting that the
number of photons is related to the energy of the field by the factor $1/\hbar
c\vert{\bf k}\vert$ and the division by $\vert{\bf k}\vert$ is represented in
the position space by a nonlocal kernel
\begin{eqnarray}
 \int\!d^3k\,e^{i{\bf k}\cdot({\bf r}-{\bf r}')}\frac{1}{\vert{\bf k}\vert} =
 \frac{1}{4\pi^2\vert{\bf r} - {\bf r}'\vert^2}.
\end{eqnarray}
The functional (\ref{phot_n}) that made its appearance in the exponent of the
Wigner functional for the vacuum state has also been used as a natural norm for
the electromagnetic field and it is intimately connected with the concept of the
photon wave function \cite{ibb5,ibb6}. Despite its ``nonrelativistic''
appearance $N\left[{\bf B},{\bf D}\right]$ is invariant not only under all
Poincar\'e transformations but also, as shown by Gross \cite{gross}, under the
full conformal group. The formula (\ref{phot_n}) teaches us an important lesson
on the subject of relativistic invariance that appearances are sometimes
misleading.

A general Wigner functional (\ref{wig_em}) is, of course, not invariant under
the Poincar\'e transformations, but it does have well defined transformation
properties that follow from its physical meaning. As a (pseudo) probability
distribution in the phase space, the Wigner function and the Wigner functional
must transform as scalars, because the probability is a scalar and the
phase-space volume is an invariant. This requirement is even more obvious for
the squeezed states, when the Wigner functional is a true probability
distribution. Thus, under a Poincar\'e transformation, the Wigner functional
transforms as follows
\begin{eqnarray}
 W'\left[{\bf B}',{\bf D}'\right] = W\left[{\bf B},{\bf D}\right].
\end{eqnarray}
The transformed functional evaluated for the transformed values of its arguments
is equal to the original functional for the original values of the arguments.
The relativistic invariance of the theory means that there exists a set of
generators of the Poincar\'e transformations that obey the commutation relations
characteristic of the Poincar\'e group. These generators are identified with the
corresponding physical quantities: energy, momentum, angular momentum, and the
center of mass. In our case these generators are given by the standard
expressions for the electromagnetic field and they act on the Wigner functional
through the Poisson bracket relations. For example, the energy of the
electromagnetic field generates the time translation and that leads to the time
evolution equation for the Wigner functional
\begin{eqnarray}
&&\partial_t W\left[{\bf B},{\bf D}\vert t\right] = \{\frac{1}{2}\int\!\!d^3r\,
 \bigl[\frac{{\bf D}^2({\bf r})}{\epsilon({\bf r})} + \frac{{\bf B}^2({\bf
r})}{\mu({\bf r})}\bigr], W\left[{\bf B},{\bf D}\vert t\right]\}\\ &&= -
\int\!d^3r\left(\nabla\times\frac{{\bf D}({\bf r})}{\epsilon({\bf r})}\!\cdot\!
 \frac{\delta}{\delta{\bf B}({\bf r})} - \nabla\times\frac{{\bf B}({\bf
 r})}{\mu({\bf r})}\!\cdot\! \frac{\delta}{\delta{\bf D}({\bf r})}\right)
 W\left[{\bf B},{\bf D}\vert t\right]\nonumber
\end{eqnarray}
that is the counterpart of  (\ref{time_wig}). The complete analysis of
relativistic invariance along these lines has been carried out in detail in
Refs. \cite{qed,bbht}. The second reference is particularly relevant since it
deals with the classical statistical theory of the electromagnetic field and all
the formal properties of the Wigner functional are the same as those of the
classical distribution function.

\section*{Acknowledgements}

I would like to thank Wolfgang Schleich for the hospitality at the Ab\-teil\-ung
f\"ur Quantenphysik of the University of Ulm where part of this research has
been carried out.
\label{pag:lab}

\end{document}